\documentclass[prc,aps,twocolumn,showpacs,amssymb,superscriptaddress,float]{revtex4}

\usepackage{amsmath}
\usepackage{graphicx}
\usepackage{dcolumn}
\usepackage{float}

\begin{document}

\title{ $g_{9/2}$ nuclei and neutron-proton interaction}

\author{L. Coraggio}
\affiliation{Istituto Nazionale di Fisica Nucleare, 
Complesso Universitario di Monte S. Angelo, I-80126 Napoli,
Italy}
\author{A. Covello}
\affiliation{Istituto Nazionale di Fisica Nucleare, 
Complesso Universitario di Monte S. Angelo, I-80126 Napoli,
Italy}
\affiliation{Dipartimento di Scienze Fisiche, Universit\`a
di Napoli Federico II,
Complesso Universitario di Monte S. Angelo,  I-80126 Napoli,
Italy}
\author{A. Gargano}
\affiliation{Istituto Nazionale di Fisica Nucleare, 
Complesso Universitario di Monte S. Angelo, I-80126 Napoli,
Italy}
\author{N. Itaco} 
\affiliation{Istituto Nazionale di Fisica Nucleare, 
Complesso Universitario di Monte S. Angelo, I-80126 Napoli,
Italy}
\affiliation{Dipartimento di Scienze Fisiche, Universit\`a
di Napoli Federico II,
Complesso Universitario di Monte S. Angelo,  I-80126 Napoli,
Italy}

\date{\today}

\begin{abstract}
We have performed shell-model calculations for nuclei below $^{100}$Sn, focusing 
attention on the two $N=Z$ nuclei $^{96}$Cd and
$^{92}$Pd,  the latter having been recently the subject of great experimental and theoretical 
interest. We have considered nuclei for which the $0g_{9/2}$ orbit plays a dominant 
role and employed a realistic low-momentum two-body effective interaction  derived from the CD-Bonn nucleon-nucleon potential. 
This implies that  no phenomenological input enters our effective Hamiltonian. The calculated results for $^{92}$Pd are in very 
good agreement with the available experimental data, which gives confidence in 
our predictions for  $^{96}$Cd. An analysis of the wave functions of both
 $^{96}$Cd and $^{92}$Pd is performed to investigate the role of the isoscalar  spin-aligned coupling.

\end{abstract} 

\maketitle

\section{Introduction}

In a recent work \cite{Cederwall11} three excited states in the $N=Z$ nucleus $^{92}$Pd, lying quite far from the  stability line, were observed. This is a remarkable achievement since the $N=Z$ nuclei play a special role in understanding the nuclear effective interaction, in particular the interplay and competition between the  isovector and isoscalar components. In fact,  in this case neutrons and protons occupy the same orbitals, which gives rise to a large overlap of their wave functions. In this context, the main question, which is still a matter of debate \cite{Satula97,Macchiavelli00}, is the existence of strongly correlated $T=0$ $np$ pairs, similarly to the well-known case of neutron and proton pairs coupled to $J=0$ and $T=1$. 

In Ref. \cite{Cederwall11} it was pointed  out that the main feature of the measured levels of $^{92}$Pd is their approximate equidistance. An interpretation was given of this feature in terms of a shell-model calculation with an empirical Hamiltonian. This 
reproduces  very well the experimental levels and yields wave functions built manly from  $J=9$ $np$ pairs, which
has been seen  as evidence for a spin-aligned $np$ paired phase.

The results of \cite{Cederwall11} have immediately attracted much attention and the role of the isoscalar $np$ pairs  in the low-energy structure of the $N=Z$ nuclei close to doubly magic $^{100}$Sn has been investigated in some very recent theoretical 
papers \cite{Zerguine11,Qi11,Xu12}.
These studies confirm substantially the dominance of $J=9$, $T=0$ pairs in the low-lying 
yrast states of the $N=Z$ nuclei with four, six and eight holes below $^{100}$Sn. 

The relevance of the isoscalar 
component of the $np$ interaction has been stressed  in a subsequent work~\cite{Singh11}, where the
$16^+$  ``spin-gap''  isomer  in $^{96}$Cd has been identified and its origin explained as due to the strong influence of this component.

Another interesting outcome of the above works is that in all nuclei above  $^{88}$Ru the low-lying yrast states can be essentially described in terms of the single $0g_{9/2}$ shell. This situation is of course reminiscent of the so-called $f_{7/2}$ nuclei that have been the subject of a large number of theoretical studies ever since the seminal paper by McCullen, Bayman and Zamick \cite{McCullen64}.  It seems therefore appropriate to call these special nuclei below $^{100}$Sn ``$g_{9/2}$ nuclei''.

Some ten years ago, we performed \cite{Coraggio00} shell-model calculations for $N=50$ nuclei immediately below $^{100}$Sn employing a realistic effective interaction derived from the Bonn-A nucleon-nucleon ($NN$) potential by means of a $G$-matrix formalism.  
In that work we took as model space for the valence proton  holes the four levels 
$f_{5/2}$, $p_{3/2}$, $p_{1/2}$ and $g_{9/2}$ of the 28-50 shell. Our results turned out to be in very good agreement with the available experimental data. 
Since then, however, a new paradigm for realistic shell-model calculations has been developed which consists in renormalizing the strong short-range repulsion of the bare 
$NN$ potential through the so-called $V_{\rm low-k}$ approach \cite{Bogner02}.
Furthermore, high-precision potentials have been constructed which fit the $pp$ and $np$ scattering data 
with a $\chi^2/{\rm datum} \approx 1$.
 
The exciting new findings mentioned before have stimulated us to perform modern realistic shell-model 
calculations for nuclei below $^{100}$Sn, with particular attention focused on  $^{92}$Pd 
and on the heavier $N=Z$ nucleus $^{96}$Cd,
for which theoretical predictions are likely to be verified in the not too distant future.
Based on the dominant role of the  $g_{9/2}$ orbit in the low-lying states of the nuclei considered 
in the present study, we have restricted our calculations to  the single $g_{9/2}$ shell. 
This  permits a more transparent analysis of the structure of the wave functions, especially for $^{96}$Cd, 
in terms of either the $[nn] \otimes [pp]$  or $[np] \otimes [np]$ coupling schemes. The comparison 
between these two approaches is 
instrumental to  understand the role of the $J=9$ $np$ pairs.
It may also be worth recalling that within the $fpg$ space the choice of the single-hole 
energies is not an easy task \cite{Coraggio00}, since there is no spectroscopic information on the 
single-hole nuclei $^{99}$In and $^{99}$Sn.

In Sec. II  we focus attention on the  $g_{9/2}$ effective interaction employed in our calculations. Results for $^{96}$Cd and
$^{92}$Pd are presented in Sec. III, where we also give  a detailed analysis of the structure of wave functions.
Sec. IV provides a summary and concluding remarks.

\section{$g_{9/2}$ effective interaction}

We assume $^{100}$Sn as a closed core and let the neutron and proton holes 
move in the single $g_{9/2}$ orbit. Our two-body effective interaction is derived within the framework of 
the time-dependent degenerate linked-diagram perturbation theory \cite{Coraggio09} starting from the high-precision 
CD-Bonn $NN$ potential \cite{Machleidt01}.
This potential, that  as all modern $NN$ potentials  contains high-momentum nonperturbative modes, is 
renormalized by constructing a low-momentum potential $V_{\rm low-k}$. This is achieved by integrating
out the high momentum modes of the bare potential down to a cutoff momentum $\Lambda$ = 2.1 fm$^{-1}$. Then the smooth  
$V_{\rm low-k}$ potential plus the Coulomb force for protons is used to calculate the two-body matrix elements
of the effective interaction by means of the ${\hat Q}$-box folded-diagram expansion \cite{Coraggio09}, with 
the ${\hat Q}$-box including all diagrams up to third order in $V_{\rm low-k}$. These diagrams are computed 
using
the harmonic oscillator basis and considering intermediate states composed of all possible hole states
and particle states restricted to the six  proton and neutron
shells above the Fermi surface.
The oscillator parameter $\hbar \omega$ is 8.55 MeV, as obtained from the expression
$\hbar \omega$ = 45$A^{-1/3}-25A^{-2/3}$ for $A=100$. The shell-model calculations have been performed by the
NUSHELLX code~\cite{NUSHELLX}.

\begin{table}[H]
\caption{Proton-proton, neutron-neutron, and proton-neutron  matrix elements of $V_{\rm eff}$ in the $g_{9/2}$ orbit (in MeV).} 
\begin{ruledtabular}
\begin{tabular}{llcccc}
J &  T & $ pp$ &  $nn$ & $np$ \\
\colrule
0 & 1 &-1.836 & -2.224 & -2.317 \\
1 & 0 & &  & -1.488 \\
2 & 1 &-0.353 & -0.662 & -0.667 \\
3 & 0 & &  & -0.440 \\
4 & 1 & 0.171 & -0.088 & -0.100 \\
5 & 0 & &  & -0.271 \\
6 & 1 & 0.317 & 0.083 &  0.066 \\
7 & 0 & &  & -0.404 \\
8 & 1 & 0.459 & 0.221 & 0.210 \\
9 & 0 & &  & -1.402 \\
\end{tabular}
\end{ruledtabular}
\label{tbme}
\end{table}

We show in Table~\ref{tbme} the two-body 
matrix elements of the effective interaction. Owing to the Coulomb force, 
there is no isospin symmetry, the $nn$ matrix elements being more attractive than the $pp$ 
ones by about 250-350  keV, which agrees quite well with the results of 
previous works \cite{Serduke76,Gross76} where the effective interaction was determined by a least-squares fit to experimental energies. As regards 
the $T=1$ $np$ matrix elements, they differ only 
by at most 100 keV from the $nn$ ones. This reflects the fact that for $np$ and $nn$ valence 
holes/particles 
the Coulomb force acts only through $\hat{Q}$-box diagrams starting at second and third 
order, respectively. 
From Table~\ref{tbme} we also see that the most 
attractive matrix elements correspond to the $J=0$, $J=1$, and $J=9$ states, the $J=0$ 
matrix element being  the largest one while the other two have about the same magnitude.
This dominance of the $J=0$ matrix element  stems from the fact that
our interaction is derived for  the single $g_{9/2}$ orbit. The same feature is shown  by the
three different $g_{9/2}$ interactions 
of Ref.~\cite{Zerguine11}
while this is not the case for interactions defined in the $fpg$ space, as for instance that developed in~\cite{Honma09}.

\begin{figure}[H]
\begin{center}
\includegraphics[scale=0.6,angle=0]{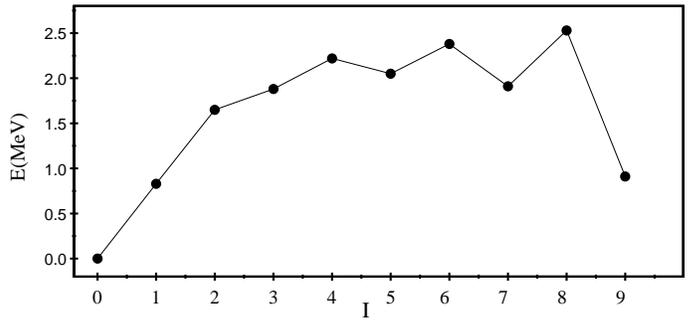}
\caption{Calculated proton hole-neutron hole   
multiplet in  $^{98}$In.}
\label{in}
\end{center}
\end{figure}

The $np$ matrix elements, relative to the $J=0$  energy, correspond to the excitation spectrum of 
$^{98}$In, which we find interesting to show in Fig.~\ref{in}. This has  unfortunately 
no experimental counterpart, so cannot be used  to test our interaction. 
We see, however,  that
the $np$ multiplet exhibits a downward parabolic behavior, which is typical of
nuclei with one proton-one neutron valence particles or holes, as for instance 
$^{42}$Sc or $^{54}$Co with two
 particles and holes, respectively,  in the $f_{7/2}$ orbit. The observed multiplet in these nuclei shows
the same behavior as that we have found for $^{98}$In, the only difference being a larger dispersion in the 
energy values which is due to the greater  attractiveness of the $np$ interaction in lighter 
systems. 

\begin{figure}[H]
\begin{center}
\includegraphics[scale=0.6,angle=0]{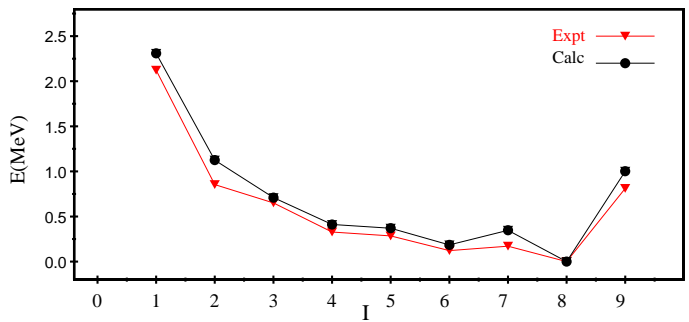}
\caption{(Color on line) Experimental and calculated proton hole-neutron  
particle multiplet in  $^{90}$Nb.}
\label{nb}
\end{center}
\end{figure}

For a test of our $np$ matrix elements, we compare in Fig.~\ref{nb} the observed
multiplet in $^{90}$Nb \cite{NNDC} with the calculated one. 
The latter is, in fact, directly
related to the matrix elements  of the hole-hole $np$ interaction 
through the Pandya transformation \cite{Pandya}. Note that several $1^+$ states have been observed
in $^{90}$Nb at low excitation energy. Following the suggestion of Ref. \cite{Zerguine11}, we report
the fifth  one  at 2.126 MeV  while exclude the observed $0^+$ state at 5.008 MeV, which is 
too high in energy to make trustable our interpretation  in terms of the single $g_{9/2}$ model.
Fig.~\ref{nb} shows that the agreement between theory and experiment 
is very good, the calculated energies overestimating the experimental values by at most 150 keV in the case of 
the $2^+$ state. The latter, however, is likely to be admixed with configurations outside our model space.

\begin{figure}[H]
\begin{center}
\includegraphics[scale=0.4,angle=0]{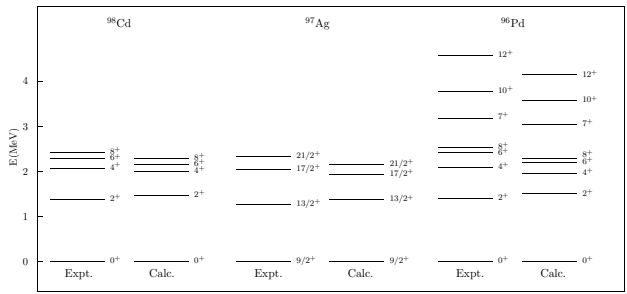}
\caption{Experimental and calculated spectra of the  $N=50$ isotones
  $^{98}$Cd, $^{97}$Ag, and $^{96}$Pd.}
\label{n50} 
\end{center}
\end{figure}

We now focus attention on the $pp$ matrix elements by
comparing the spectra of the three $N=50$ isotones $^{98}$Cd, $^{97}$Ag,
and  $^{96}$Pd with the experimental ones. We include all observed levels for $^{98}$Cd while 
positive-parity  yrast levels below 4.5 MeV for the two latter nuclei. This is done in Fig.~\ref{n50},
where we see that the agreement between theory and experiment is quite satisfactory up to about
3.7 MeV. 
We overestimate the energy of the first excited state in
the three nuclei by less than 100 keV while the energies of all
other levels, except the $12^+$ in  $^{96}$Pd which is predicted at more than 400 keV below the experimental one,
are underestimated by an amount which  
increases when the mass number decreases reaching at most 250 keV.
In concluding this section, it is worth noting that the  accuracy of the present $g_{9/2}$
calculation is similar
to that of \cite{Cederwall11}, where the spectrum of  $^{96}$Pd  up to the
$10^+$ state was calculated in the $fpg$ model space.

\section{Results}

\begin{figure}[H]
\begin{center}
\includegraphics[scale=0.6,angle=0]{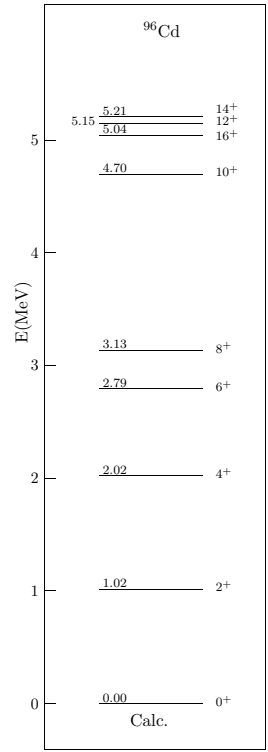}
\caption{Calculated spectrum of  $^{96}$Cd.}
\label{96Cd}
\end{center}
\end{figure}

\begin{figure}[H]
\begin{center}
\includegraphics[scale=0.5,angle=0]{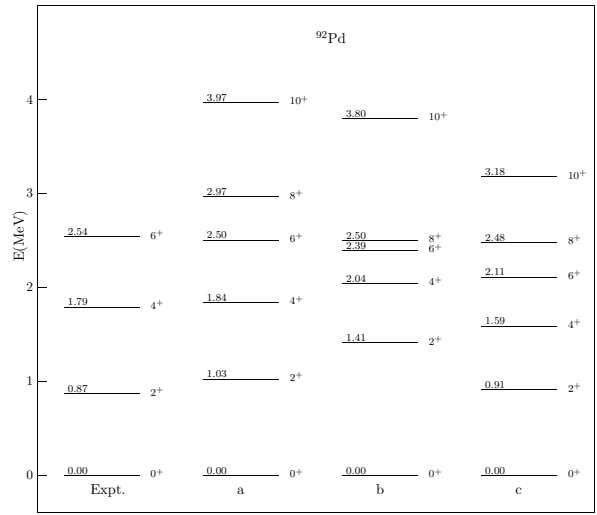}
\caption{Experimental spectrum of $^{92}$Pd compared with the results
  of calculations with: a) all matrix elements, see Table~\ref{tbme};
  b) $nn$, $pp$, $np$ $T=1$ matrix elements; c) $nn$, $pp$, $np$ $T=0$
  matrix elements.}
\label{92Pd}
\end{center}
\end{figure}

The results obtained for $^{90}$Nb and $N=50$ isotones have given us  confidence in our effective interaction, 
at least as regards 
its predictive power for low energy states. We have then  performed
calculations  for the two $N=Z$ nuclei $^{96}$Cd and $^{92}$Pd. The calculated  
spectra are reported in Figs.~\ref{96Cd} and~\ref{92Pd}a, 
respectively, together with a comparison with the experimental data of \cite{Cederwall11} for $^{92}$Pd.
In view of the recent experimental finding \cite{Singh11} we have reported yrast states in  $^{96}$Cd up to
$I^{\pi}=16^+$, while for $^{92}$Pd we have included two more states with respect to those identified in \cite{Cederwall11}.

The three observed states in $^{92}$Pd  are very well reproduced by the theory. Of course, one may not
expect the same kind of agreement for high-energy states.
It is worth mentioning, however, that our results account for the isomeric nature of the
$16^+$ state identified in $^{96}$Cd. More precisely, we find 
that the location of this state below the $I^{\pi}=12^+$ and $14^+$ states can 
be traced to the isoscalar $np$ component of the interaction, in agreement with the results of ~\cite{Singh11}. 
In particular, we have verified that a strong attractive  $J=9$ matrix element is essential 
to make the $16^+$ state isomeric.
This is related to the fact that for the $12^+$, $14^+$,  and $16^+$ states the average number 
of isoscalar $J=9$ pairs (see Sec. III.B for the definition of this quantity) increases with
angular momentum,  reaching its  maximum value for the $16^+$ state. This is, in fact, 2.5 to be 
compared with 2.2 and 2.0 for the $14^+$ and $12^+$ states, respectively.
Therefore a sufficiently attractive $J=9$ matrix element pushes the $16^+$ state 
below the two lower-spin states.
Relevant to this discussion is the fact that the structure of the wave functions for the 
$12^+$, $14^+$,  and $16^+$ states, and consequently 
the average number of isoscalar $J=9$ pairs,  does not depend on the size of the $J=9$ matrix element.

Let us now focus on states up to  $I^{\pi}=10^+$ for $^{96}$Cd and $^{92}$Pd. For both nuclei,
we find an almost regularly spaced
level sequence up to $I^{\pi}=6^+$, with a slightly reduced $6^{+}-4^{+}$ spacing.
Then, the   $8^{+}-6^{+}$  spacing  becomes even smaller while the  $10^{+}-8^{+}$ one
increases considerably.
These features, which up to the $6^+$ state find a  correspondence  in the experimental data 
for $^{92}$Pd, are only slightly  more pronounced in $^{96}$Cd.
This similarity may be seen  as an indication that the same correlations come into play in their low-energy spectra.
We shall discuss in more detail the results for both nuclei in the two next 
subsections.

\subsection{ $^{96}$Cd}

We start  with $^{96}$Cd, for which 
a simple and clear analysis  of the wave functions can be performed using the  orthogonal basis
formed by products of  $nn$ and $pp$ states.  The structure of the wave functions in terms of 
this basis set is shown in Table~\ref{v0cd}, while  their  overlaps with  
the $[(np)9 (np)9]_I$ state
are given in Table~\ref{j9cd}a.

\begin{table}[H]
\caption{Overlap of the calculated $I^{\pi}=0^{+}$, $2^{+}$, $4^{+}$, $6^{+}$, $8^{+}$, and   $10^{+}$
yrast states in $^{96}$Cd with the $[(nn)J_{n} (pp)J_{p}]_{I}$ states, expressed in 
percentage. Only components with 
percentage $>$ 10 are reported.} 
\begin{ruledtabular}
\begin{tabular}{lcccccccccc}
& \multicolumn {10} {c} {($J_{n}, J_{p}$)}  \\
 \cline{2-11}   
$I^{\pi}$ & (0,0) & (0,J) & (J,0) & (2,2)& (2,4)& (4,2)& (2,8)& (8,2)& (4,6) & (6,4) \\
\colrule
$0^+$ &  57& &  & 30& & &\\
$2^+$ & & 34 &32 & 12 & & & \\
$4^+$ & & 29 &26&  28 & & & \\
$6^+$ & & 33 &26 & &16 &15 &\\ 
$8^+$ & & 39 & 25 & &  &  & 12\\   
$10^+$& & & & & & & 17& 15& 18&17\\
\end{tabular}
\end{ruledtabular}
\label{v0cd}
\end{table}

\begin{table}[H]
\caption{Overlap of the calculated $I^{\pi}=0^{+}$, $2^{+}$, $4^{+}$, $6^{+}$, $8^{+}$, and $10^{+}$
yrast states in $^{96}$Cd with the $[(np)9 (np)9]_{I}$ state, expressed in percentage, obtained using:
a) $V_{9}(np)$ of Tab.~\ref{tbme}; b) one-half the original value of $V_{9}(np)$; 
c) twice  the original value of $V_{9}(np)$.} 
\begin{ruledtabular}
\begin{tabular}{lccc}
$I^{\pi}$ & a & b & c \\
\colrule
$0^+$ &  90 & 82 & 96\\
$2^+$ & 97 & 94 & 99\\
$4^+$ & 85 & 73 & 94\\
$6^+$ & 48 & 27 & 84\\
$8^+$ & 6 & 3 & 27\\
$10^+$ & 46 & 13 &94 \\
\end{tabular}
\end{ruledtabular}
\label{j9cd}
\end{table}

Note that for each angular momentum $I$  an orthonormal  basis  formed by  products of two $np$-pair
vectors  can been  constructed, the  $[(np)9 (np)9]_I$ state being  one
of them. These vectors  are  simply related to the $[nn] \otimes [pp]$ 
basis through

\begin{eqnarray}
|(np){J_1}(np){J_2};I \rangle = \frac{1}{\sqrt{N_{J_{1}J_{2}}}}\sum_{J_{n}J_{p}} [\hat{J_{1}}\hat{J_{2}}\hat{J_{n}}
\hat{J_{p}}]^{1/2} \nonumber \\ 
\left \{ \begin{array}{ccc}
\frac{9}{2} & \frac{9}{2} & J_{1} \\
\frac{9}{2} & \frac{9}{2} & J_{2} \\
J_{n} &  J_{p} &         I
\end{array}
\right \}
|(nn){J_n}(pp){J_p};I \rangle ~~,
\end{eqnarray}

\noindent{where $[\hat{J}]=(2J+1)$ and $N$ denotes the normalization factor.}
From Table~\ref{v0cd} we see that the  wave functions of $^{96}$Cd expressed in terms 
of the $[nn] \otimes [pp]$ basis  are strongly  fragmented. In fact,          
large seniority-4 components  are present in the ground as well as in the
excited $I^{\pi}=2^+$, $4^+$, $6^+$, and  $8^+$  states.  More precisely, their percentage in the 
ground  state is 43\% while in the other states is not less than 34\%.  
The $10^+$ state is characterized by an admixture of different seniority-4 components, each with 
a weight not exceeding 18\%. 
On the other
hand, Table~\ref{j9cd}a shows that,  when written in terms of an $[np] \otimes [np]$ basis, 
the ground and the first  two excited states are  largely  dominated by the  
$[(np)9 (np)9]_I$ component. The weight of this component, however,  is significantly smaller for the 
other three states having the minimum value (6\%) for the $8^+$ state.
These results for  $^{96}$Cd are in line with those of Ref.~\cite{Zerguine11}.

The large fragmentation evidenced in Table~\ref{v0cd} is of course related to the $np$ 
interaction and in this connection it is 
interesting to find out what is the role of  the 
$J=9$ matrix element $V_{9}(np)$. To this end, we
have redone our calculations with two different  values of $V_{9}(np)$, namely 
increasing and reducing  $V_{9}(np)$  by a factor of 2.
A similar analysis was done in \cite{Qi11,Xu12}.

We find  that for the reduced value of $V_{9}(np)$ the $6^+$ and $8^+$ states come down in energy
getting close to the $4^+$ state. As a matter of fact, in this case  the group of  the $4^+$, $6^+$, 
and  $8^+$ levels concentrates in a small energy range 
separated by a large energy gap  from both the $2^+$ and $10^+$ states. The spectrum 
of $^{96}$Cd (up to $I^{\pi}=8^{+}$) becomes then similar to that of $^{98}$Cd, the 
wave functions being, however, still significantly seniority admixed.
The weight of the seniority-4 components is no smaller than 29\%.
On the contrary, a doubled value  of $V_{9}(np)$  
leads to almost equidistant levels, with an energy separation of about 1 MeV, the only
exception being the $8^{+}-6^+$ spacing which is  $\sim 300$ keV smaller. 
This evolution toward an equidistant-level spectrum comes along with a larger  fragmentation 
of the wave functions. The only state which has a less admixed nature  is
the $10^+$ state. 
We find that  the percentage of the $[(nn)0 (pp)0]_0$ component in the 
ground state reduces to  51\%, while  that of the  seniority-two components,
$[(nn)I (pp)0]_I$ and $[(nn)0 (pp)I]_I$, in the 
$I=2^{+}$, $4^{+}$, $6^{+}$, and $8^{+}$ states ranges from  33\% to 63\%.
These values, however, remain significantly large  showing that pairing is still in the game.

We now examine the influence of $V_{9}(np)$ on  
the dominance of  isoscalar $J=9$ pairs in the wave functions of $^{96}$Cd. To this end,  we 
have also  reported in
Table~\ref{j9cd}  the overlaps with the $[(np)9 (np)9]_I$ state obtained using one-half (b) and 
twice (c) the original  $V_{9}(np)$  value. 
We see that a larger value of $V_{9}(np)$ leads to an  increase of the overlap for all the states.
However the overlap for the $8^+$ state does not go beyond 27\%. 
Moreover, it should be noted that for the  three lowest-lying states the overlaps do not 
become significantly smaller even when $V_{9}(np)$  is reduced. This is related to
the structure of the $[(np)9 (np)9]_{I=0,2,4}$ states  in terms of the $[nn] \otimes [pp]$ basis [see~Eq.~(1)].
Therefore, as already  
pointed out  in~\cite{Zerguine11}, both dynamics and geometry are crucial to
 the presence of $J=9$ pairs
in the  wave functions of $^{96}$Cd.

\subsection{$^{92}$Pd}

The simple analysis done for $^{96}$Cd cannot be performed for $^{92}$Pd
with four  neutron and  four proton holes. 
In this case, we shall first discuss  the effects  of the $T=0$ 
and $T=1$ components of the interaction
on the calculated spectrum. This study was already performed
by Cederwall {\it et al}.~\cite{Cederwall11}, and we have found it worth verifying if
a realistic effective interaction confirms their results.

In Fig.~\ref{92Pd} the results of the full-interaction calculation (a) are compared with those 
obtained
by removing  separately the  $T=0$ (b) and $T=1$ (c) $np$ matrix elements.
We see that in case (b) the excited states up to $I^{\pi}= 8^+$ are compressed  in a smaller energy 
interval, about 1 MeV to be compared with 2 MeV of the full 
calculation. Actually, the spectrum of $^{92}$Pd 
evolves toward that of the neutron closed
shell nucleus, as it was the case for $^{96}$Cd when using a reduced value of $V_{9}(np)$.
On the other 
hand, when we exclude the $T=1$ $np$ matrix elements all the excited levels move down, but 
the  spectrum keeps the same structure as that obtained from the full calculation. 

Our findings are in line with those of Ref.~\cite{Cederwall11}, confirming the more 
relevant role of the $T=0$   versus the $T=1$ $np$ component. More specifically, we have verified that 
the addition  of the  sole $J=9$ $np$ matrix element to the interaction between identical 
particles 
is sufficient to produce a spectrum very similar to that of Fig.~\ref{92Pd}c. 
We may therefore conclude that  the structural makeup of  the $^{92}$Pd spectrum is mainly determined 
by the combined action of  $J=9$ $np$, $nn$ and $pp$ matrix elements. Needless to say,  
a quite
distorted, highly compressed spectrum would result from  ignoring the interaction between identical particles.

In this context we have tried to better understand  how different isoscalar and isovector pairs contribute to produce the spectrum of 
 Fig.~\ref{92Pd}a.
 To this end, we have used  the relation

\begin{equation}
E_{I}=\sum_{J}
[C^{I}_{J}(np)V_{J}(np)+C^{I}_{J}(pp)V_{J}(pp)+C^{I}_{J}(nn)V_{J}(nn)]
~~, 
\end{equation}

\noindent {where  the energy of  a given state  is written  in terms of the average numbers of $nn$, $pp$, and $np$ pairs, 
$C^{I}_{J}(ij)$'s, defined as $C^{I}_{J}(ij)= <\psi_{I} \, (^{92}{\rm Pd})\,|\, [(a^{\dag}_{i} \, a^{\dag}_{j})_{J} \, \times (a_{i} \, a_{j})_{J}]_{0}\, |\,\psi_{I} \, (^{92}{\rm Pd})>$.
In Eq.~(2)
the matrix elements in the three different channels appear explicitly,  
since, as mentioned in Sec. II,  our effective 
interaction includes the Coulomb force. For the
sake of simplicity,
 in the following we shall not  distinguish between 
$nn$, $pp$, and $np$ isovector pairs and  take as $V_{J}$, with $J$ even, the mean value of the 
three corresponding matrix elements.}

\begin{figure}[H]
\begin{center}
\includegraphics[scale=0.6,angle=0]{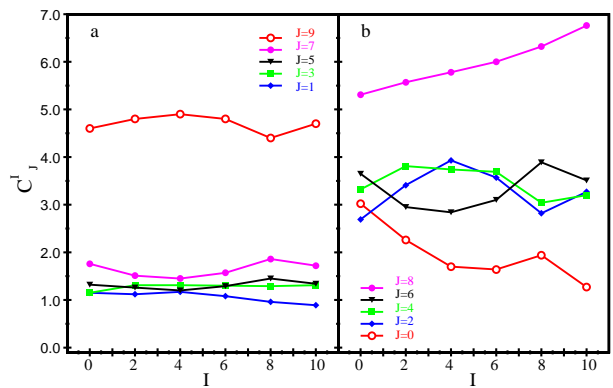}
\caption{(Color on line) Average number of the isoscalar (a) and
  isovector (b) ($g_{9/2}$)$^{2}J$ pairs, $C^{I}_{J}$,  as a function
  of the angular momentum $I$ of the yrast states in $^{92}$Pd.}
\label{numm}
\end{center}
\end{figure}

In Figs.~\ref{numm}a and  \ref{numm}b, we show
the average  number of the isoscalar and isovector pairs for the six considered yrast states. 
 We see that the curves  corresponding to $J=8$ and 9 lie 
significantly higher than the others, as was already observed in  ~\cite{Qi11}.
We draw attention here on the fact that the curves of Fig. \ref{numm}a are  almost 
flat when compared to those of  Fig. \ref{numm}b. This means [see Eq.~(2)] that 
the isovector pairs  contribute  to the level spacings 
more significantly than the isoscalar ones.
In particular, a main role is played by the  $J=0$ and 2 pairs, whose 
corresponding matrix elements are much larger than the other ones. 

\begin{table}[H]
\caption{Contributions (in MeV) of isoscalar and isovector pairs to the energy 
level spacings of $^{92}$Pd. 
 Values  $<$ 0.13  are omitted. Column $S$ gives the sum of contributions 
while column
 $\Delta E$ the energy spacing obtained from the full calculations. See text for details.} 
\begin{ruledtabular}
\begin{tabular}{lccccccc}
$I^{\pi}_{i}-I^{\pi}_{j}$ & $J=0$ & $J=2$ & $J=8$ & $J=1$ & $J=9$ & $S$ & $\Delta E$ \\
\colrule
$2^{+}-0^+$ &  1.62 & -0.40 &  &  &-0.28& 0.94& 1.03\\
$4^{+}-2^+$ & 1.19 & -0.29&    & &-0.14&0.76&0.81 \\
$6^{+}-4^+$ & 0.13 & 0.20&      & 0.13& 0.14 &0.60 & 0.66\\
$8^{+}-6^+$ & -0.64 & 0.42 &   & 0.18& 0.56 & 0.52 &0.47\\
$10^{+}-8^+$ & 1.42 &-0.26&0.13 &  &-0.42& 0.88& 1.00 \\
\end{tabular}
\end{ruledtabular}
\label{spac}
\end{table}

The contributions to the five spacings arising from the different pairs are reported in 
Table~\ref{spac}, where we only include values larger than 130 keV.
The two last columns allows to compare the calculated spacings of Fig.~\ref{92Pd}a  with those obtained by summing the 
contributions reported in the table. We see that they do not  differ significantly.
From Table~\ref{spac} it appears that the most important contributions to the energy spacings 
arise from the $J=0$ pairs, although those from $J=2$ and 9 cannot be 
ignored at all, being  particularly relevant for the $6-4$ and $8-6$ spacings. In this regard, 
it should be kept in mind that the structure of the wave functions, and consequently the specific action of
the interaction in the $J=0$ and 2 channels, is strongly influenced by the size of the $J=9$ matrix element.

In concluding this discussion, we should mention that the content of isoscalar
spin-aligned pairs in  the
wave functions of $^{92}$Pd
is still significant even when $V_{9}(np)$ is suppressed. This, as already emphasized for  $^{96}$Cd, is
related to geometrical features.

\section{Summary and Concluding remarks}

In this work, we have performed a shell-model study of $N=Z$ nuclei below $^{100}$Sn 
that can be described in terms of the single $g_{9/2}$ orbit. The effective interaction for this orbit 
has been derived from the CD-Bonn $NN$ potential without using any adjustable parameter.
This approach reproduces very well the excited states of $^{92}$Pd observed in a recent experiment, and 
therefore provides confidence in our predictions for  $^{96}$Cd and $^{98}$In.

Aside from the intrinsic interest in employing a realistic effective interaction to describe these neutron-deficient nuclei, the present work was motivated by the suggestive interpretation given in Ref.~\cite{Cederwall11} of the almost equidistant levels in the $^{92}$Pd spectrum. It was argued, in fact, that this feature may be traced to an isoscalar spin-aligned $np$ coupling scheme replacing the normal isovector $J=0$ pairing which is dominant for like valence particle nuclei.
We have thus decided to investigate the role played by the $np$ interaction
in the $J=9$, $T=0$ channel. 

As regards $^{96}$Cd, we have found that  the matrix element $V_{9}(np)$ has a direct influence on   
energies of the high-spin states $12^+$, $14^+$ and $16^+$, making the latter isomeric, while for 
the states with $I^{\pi}$ from $0^+$ to $10^+$ it also substantially affects the structure of the
wave functions. More precisely, a more attractive matrix element leads to a larger content of $J=9$ 
pairs and, for states up to $I^{\pi}=8^+$, to a larger fragmentation in terms of the 
$[nn] \otimes[pp]$ basis. It should be mentioned, however, that the large content of $J=9$ pairs 
for $I^{\pi}=0^{+}$, $2^{+}$ and $4^{+}$ arises also from  the significant overlap of  the $[(np)9 (np)9]_I$ 
component with $[(nn)0 (pp)I]_I$ and $[(nn)I (pp)0]_I$. 

In agreement with previous papers, we have found  that the $J=9$ matrix element plays an important role in determining the low-energy spectrum of both $^{96}$Cd and $^{92}$Pd. However, as shown in some detail for $^{92}$Pd, it does not contribute to the energy spacings directly, but rather by changing the weights of the isovector contributions through the induced fragmentation. So to say, the pairing force pushed out the door comes back through the window.

In summary, we confirm the relevant role  of the isoscalar spin-aligned coupling 
in the low-energy states of $^{92}$Pd and $^{92}$Cd. Based on our study and the presently available data 
for $^{92}$Pd, we feel, however, that one can hardly speak of a new phase of nuclear matter similar to the well-known one induced by the strong pairing correlations between identical particles.

\end{document}